\documentclass[leqno]{article}
\usepackage{fullpage}
\usepackage{amsthm}
\usepackage{amsmath,amssymb}
\usepackage{color}
\usepackage[colorlinks=true,urlcolor=webblue,linkcolor=webgreen,filecolor=webblue,citecolor=webgreen,pdfpagemode=UseOutlines,pdfstartview=FitH,pdfpagelayout=OneColumn,bookmarks=true]{hyperref}

\hypersetup{
  pdftitle=The Power of Strong Fourier Sampling: Quantum Algorithms for 
  Affine Groups and Hidden Shifts,
  pdfauthor=Cristopher Moore and Daniel Rockmore and Alexander Russell and Leonard J. Schulman}

\definecolor{webgreen}{rgb}{0,.5,0}
\definecolor{webblue}{rgb}{0,0,.5}

\newtheorem{theorem}{Theorem}
\newtheorem{lemma}{Lemma}

\newcommand{\remove}[1]{}

\newcommand{\ket}[1]{\left| #1 \right\rangle}
\newcommand{\bra}[1]{\left\langle #1 \right|}

\newcommand{\C}{{\mathbb C}}
\newcommand{\F}{{\mathbb F}}
\newcommand{\Z}{{\mathbb Z}}
\newcommand{\ord}{O}
\newcommand{\e}{{\rm e}}
\newcommand{\eps}{\epsilon}
\newcommand{\poly}{{\rm poly}}
\newcommand{\polylog}{{\rm polylog}}
\newcommand{\re}{{\rm Re}\,}
\newcommand{\U}{\textrm{U}}
\newcommand{\id}{\mathbf{1}}
\newcommand{\tr}{\textbf{tr}}

\newcommand{\cg}{\frac{1}{\sqrt{|G|}}}

\newcommand{\ve}{\left( \! \begin{array}{c}}
\newcommand{\ctor}{\end{array} \! \right)}
\newcommand{\mat}{\left( \! \begin{array}{rr}}
\newcommand{\rix}{\end{array} \! \right)}
\newcommand{\norm}[1]{\left\| #1 \right\|}
\newcommand{\abs}[1]{\left| #1 \right|}

\newcommand{\rank}{\textbf{rk}\;}

\newcommand{\proc}{Proc.\ }

\begin{document}

\title{The Power of Strong Fourier Sampling:
Quantum Algorithms for Affine Groups and Hidden Shifts}

\author{Cristopher Moore\\University of New Mexico\\\textsf{moore@cs.unm.edu}
\and Daniel Rockmore\\ Dartmouth College\\\textsf{rockmore@cs.dartmouth.edu}
 \and
Alexander Russell\\ University of Connecticut\\\textsf{acr@cse.uconn.edu}
\and Leonard J. Schulman\\ California Institute of Technology\\\textsf{schulman@caltech.edu}}

\date{}

\maketitle

\begin{abstract}
  Many quantum algorithms, including Shor's celebrated factoring and
  discrete log algorithms, proceed by reduction to a \emph{hidden
    subgroup problem}, in which an unknown subgroup $H$ of a group $G$
  must be determined from a quantum state $\psi$ over $G$ that is
  uniformly supported on a left coset of $H$.  These hidden subgroup
  problems are typically solved by \emph{Fourier sampling}: the
  quantum Fourier transform of $\psi$ is computed and measured. When the
  underlying group is nonabelian, two important variants of the
  Fourier sampling paradigm have been identified: the \emph{weak
    standard method}, where only representation \emph{names} are
  measured, and the \emph{strong standard method}, where full
  measurement (i.e., the row and column of the representation, in a
  suitably chosen basis, as well as its name) occurs. It has remained
  open whether the strong standard method is indeed stronger, that is,
  whether there are hidden subgroups that can be reconstructed via the
  strong method but \emph{not} by the weak, or any other known,
  method.

  In this article, we settle this question in the affirmative. We show
  that hidden subgroups $H$ of the $q$-hedral groups, i.e., semidirect
  products $\Z_q \ltimes \Z_p$ where $q \mid (p-1)$, and in particular
  the affine groups $A_p$, can be information-theoretically
  reconstructed using the strong standard method.  Moreover, if $|H| =
  p/ \polylog(p)$, these subgroups can be fully reconstructed with a
  polynomial amount of quantum and classical computation.

  We compare our algorithms to two weaker methods that have been
  discussed in the literature---the ``forgetful'' abelian method, and
  measuring in a random basis---and show that both of these are weaker
  than the strong standard method with a chosen basis.  Thus, at least
  for some families of groups, it is crucial to use the full power of
  representation theory and nonabelian Fourier analysis: namely, to
  measure the high-dimensional representations in an \emph{adapted
    basis} that respects the group's subgroup structure.

  We apply our algorithm for the hidden subgroup problem to new families of
  cryptographically motivated \emph{hidden shift problems},
  generalizing work of van Dam, Hallgren and Ip on shifts of
  multiplicative characters. Finally, we close by proving a simple
  closure property for the class of groups over which the hidden
  subgroup problem can be solved efficiently.
\end{abstract}

\section{The Hidden Subgroup Problem}

One of the principal quantum algorithmic paradigms is the use of the abelian 
Fourier transform to discover a function's hidden periodicities.  In the 
examples relevant to quantum computing, an oracle function $f$ defined on an 
abelian group $G$ has ``hidden periodicity'' if there is a ``hidden'' subgroup 
$H$ of $G$ so that $f$ is precisely invariant under translation by $H$ or, 
equivalently, $f$ is constant on the cosets of $H$ and takes distinct values on 
distinct cosets. The \emph{hidden subgroup problem} is the problem of 
determining the subgroup $H$ from such a function. Algorithms for these problems 
typically adopt the approach detailed below, called \emph{Fourier sampling} 
\cite{BernsteinV93}:

\begin{description}
\item[Step 1.] Prepare two registers, the first in a uniform superposition over 
the elements of a group $G$ and the second with the value zero, yielding the 
state 
$$
\psi_1 = \cg \sum_{g \in G} \ket{g}  \otimes \ket{0} \enspace.
$$
\item[Step 2.] Calculate (or if it is an oracle, query) the function $f$ defined on 
$G$ and XOR it with the second register.  This entangles the two registers and 
results in the state
$$
\psi_2 = \cg \sum_{g \in G} \ket{g} \otimes \ket{f(g)} \enspace.
$$
\item[Step 3.] Measure the second register.  This produces a uniform 
superposition over one of $f$'s level sets, i.e., the set of group elements $g$ 
for which $f(g)$ takes the measured value $f_0$.  As the level sets of $f$ are 
the cosets of $H$, this puts the first register in a uniform distribution over 
superpositions on one of those cosets, namely $cH$ where $f(c)=f_0$ for some 
$f_0$.  Moreover, it disentangles the two registers, resulting in the state 
$\psi_3 \otimes |f_0\rangle$ where 
$$
\psi_3 = \frac{1}{\sqrt{|H|}} \ket{cH} = \frac{1}{\sqrt{|H|}} \; \sum_{h \in H} \ket{ch}  \enspace.
$$
Alternately, since the value $f_0$ we observe has no bearing on the algorithm, 
we can use the formulation in which the 
environment, rather than the user, measures $f$.  In that case, tracing over $f$ 
yields a mixed state with density matrix 
\[ \frac{1}{[G:H]} \sum_{f} \ket{\psi_3} \bra{\psi_3} 
= \frac{1}{|G|} \sum_c \ket{cH} \bra{cH} \enspace , 
\] 
i.e., a classical mixture consisting of one pure state 
$\psi_3$ for each coset.  Kuperberg refers to this as the {\em coherent} 
hidden subgroup problem~\cite{Kuperberg03}.
  
\item[Step 4.] Carry out the quantum Fourier transform on $\psi_3$ and measure the
result.
\end{description}

For example, in Simon's algorithm \cite{Simon97}, 
the ``ambient'' group $G$ over which the Fourier transform is performed 
is $\Z_2^n$, $f$ is an oracle with the promise that $f(x)=f(x+y)$ for some $y$, 
and $H=\{0,y\}$ is a subgroup of order $2$. In Shor's factoring algorithm 
\cite{Shor97} $G$ is the group $\Z_n^*$ where $n$ is the number we wish to 
factor, $f(x) = r^x \bmod n$ for a random $r < n$, and $H$ is the subgroup of 
$\Z_n^*$ of index order$(r)$. (However, since $|\Z_n^*|$ is unknown, Shor's 
algorithm actually performs the transform over $\Z_q$ where $q$ is polynomially 
bounded by $n$; see \cite{Shor97} or \cite{HalesH99,HalesH00}.)

These are all abelian instances of the \emph{hidden subgroup problem} (HSP). 
Interest in \emph{nonabelian} versions of the HSP evolved from the relation to 
the elusive \textsc{Graph Automorphism} problem: it would be sufficient to solve 
efficiently the HSP over the symmetric group $S_n$ in order to have an 
efficient quantum algorithm for graph automorphism (see, e.g., 
Jozsa~\cite{Jozsa00} for a review).  This was the impetus behind the development 
of the first nonabelian quantum Fourier transform~\cite{beals} and is,
in part, the reason that the nonabelian HSP has remained such an active 
area of research in quantum algorithms.

In general, we will say that the HSP for a family of groups $G$ has a 
\emph{Fourier sampling} algorithm if a procedure similar to that outlined above 
works.  Specifically, the algorithm prepares a superposition of the form
$$
\frac{1}{\sqrt{|H|}} \sum_{h \in H} |ch\rangle,
$$
over a random coset $cH$ of the hidden subgroup $H$, 
computes the (quantum) Fourier transform of this state, and measures the result.  
After a polynomial number of such trials, a polynomial amount of classical 
computation, and, perhaps, a polynomial number of classical queries to the 
function $h$ to confirm the result, the algorithm produces a set of generators 
for the subgroup $H$ with high probability. 

When $G$ is abelian, measuring a state's Fourier transform has a clear meaning: 
one observes the frequency $\chi$ with probability equal to the squared 
magnitude of the transform at that frequency. In the case where $G$ is a 
\emph{nonabelian} group, however, it is necessary to select bases for each 
representation of $G$ to perform full measurement.  (We explain this in more 
detail below.)  The subject of this article is the relationship between this 
choice of basis and the information gleaned from the measurement: are some bases 
more useful for computation than others?  

Since we are typically interested in exponentially large groups, we will take 
the size of our input to be $n = \log |G|$.  Throughout, ``polynomial'' means 
polynomial in $n$, and thus polylogarithmic in $|G|$.

\subsection{Nonabelian Hidden Subgroup Problems}

Although a number of interesting results have been obtained on the nonabelian HSP, 
the groups for which efficient solutions are known remain woefully
few. On the positive side, Roetteler and Beth~\cite{RoettelerB98} give
an algorithm for the wreath product $\Z_2^k\; \wr\; \Z_2$.  Ivanyos, Magniez, and 
Santha~\cite{IvanyosMS01} extend this to the more general case of semidirect 
products $K \ltimes \Z_2^k$ where $K$ is of polynomial size, and also give an 
algorithm for groups whose commutator subgroup is of polynomial size.  Friedl, 
Ivanyos, Magniez, Santha and Sen solve a problem they call Hidden Translation, 
and thus generalize this further to what they call ``smoothly solvable'' groups: 
these are solvable groups whose derived series is of constant length and whose 
abelian factors are each the direct product of an abelian group of bounded 
exponent and one of polynomial size~\cite{FriedlIMSS02}. (See also 
Section~\ref{sec:closure}.)

In another vein, Ettinger and H{\o}yer~\cite{EttingerH98} show that the HSP is 
solvable for the dihedral groups in an \emph{information-theoretic} sense; 
namely, a polynomial number of quantum queries to the function oracle gives 
enough information to reconstruct the subgroup, but the best known 
reconstruction algorithm takes exponential time.  More generally, Ettinger, 
H{\o}yer and Knill~\cite{EttingerHK04} show that for \emph{arbitrary} groups the 
HSP can be solved information-theoretically with a finite number of quantum 
queries.  However, their algorithm calls for a quantum measurement for each 
possible subgroup, and since there might be $|G|^{\Omega(\log |G|)}$ of these, 
it requires an exponential number of quantum operations.

Our current understanding of the HSP, then, divides group families into three 
classes.
\begin{description}
\label{classification}
\item[I.] \textbf{Fully Reconstructible.}  Subgroups of a family of groups $\{ 
G_i \}$ are \emph{fully reconstructible} if the HSP can be solved with high 
probability by a quantum circuit of size polynomial in $\log |G_i|$.
\item[II.] \textbf{Information-Theoretically Reconstructible.}
  Subgroups of a family of groups $\{ G_i \}$ are
  \emph{information-theoretically reconstructible} if the solution to
  the HSP for $G_i$ is determined information-theoretically by the
  fully measured result of a quantum circuit of size polynomial in
  $\log |G_i|$.
\item[III.] \textbf{Quantum Information-Theoretically
    Reconstructible.}  Subgroups of a family of groups $\{ G_i \}$ are
  \emph{quantum information-theoretically reconstructible} if the
  solution to the HSP for $G_i$ is determined by the quantum state
  resulting from a quantum circuit of polynomial size in $\log |G_i|$,
  in the sense that there exists a positive operator-valued
  measurement (POVM) that yields the subgroup $H$ with constant
  probability, but where it may or may not be possible to carry out
  this POVM with a quantum circuit of polynomial size.
\end{description}
In each case, the quantum circuit has oracle access to a function $f : G \to S$, 
for some set $S$, with the property that $f$ is constant on each left coset of a 
subgroup $H$, and distinct on distinct cosets.

In this language, then, subgroups of abelian groups are fully reconstructible, 
while the result of \cite{EttingerHK04} shows that subgroups of arbitrary groups 
are quantum information-theoretically reconstructible.  The other work cited 
above has labored to place specific families of nonabelian groups into the more 
algorithmically meaningful classes I and II.


\subsection{Nonabelian Fourier transforms}

In this section we give a brief review of nonabelian Fourier analysis, but only 
to the extent needed to set down notation.  We refer the reader 
to~\cite{Serre77} for a more complete exposition. 

Fourier analysis over a finite abelian group $A$ expresses a 
function $\phi: A \to \C$ as a linear combination of homomorphisms $\chi: A \to 
\C$.  If $A = \Z_p$, for example, these are the familiar basis functions 
$\chi_t: z \mapsto \omega_p^{tz}$,
where $\omega_p$ denotes the $p$th root of unity $\e^{2 \pi i/p}$.  Any function 
$\phi : A \to \C$ can be uniquely expressed as a linear combination of these 
$\chi_t$, and this change of basis is the Fourier transform. 

When $G$ is a nonabelian group, however, this same procedure cannot work: in 
particular, there are not enough homomorphisms of $G$ into $\C$ to 
span the space of all $\C$-valued functions on $G$.  To define a sufficient 
basis, the representation theory of finite groups considers more general 
functions,  namely homomorphisms from $G$ into groups of unitary matrices.

A \emph{representation} of a finite group $G$ is a homomorphism $\rho: G \to 
\U(d)$, where $\U(d)$ denotes the group of unitary $d \times d$ matrices (with 
entries from $\C$); the dimension $d = d_\rho$ is referred to as the 
\emph{dimension} of $\rho$. If $\rho: G \to \U(d)$ is a representation, a 
subspace $W$ of $\C^d$ is said to be \emph{invariant} if $\rho(g)(W) \subset W$ 
for all $g$.  A representation is said to be \emph{irreducible} if the only 
invariant subspaces are the trivial subspace $\C^d$ and $\{ \vec{0} \}$.

For a function $\phi: G \to \C$ and an irreducible representation $\rho$, 
$\hat{\phi}(\rho)$ denotes \emph{the Fourier transform of $\phi$ at $\rho$} and 
is defined by
\[
\hat{\phi}(\rho) = \sqrt{\frac{d_\rho}{|G|}} \,\sum_g \phi(g)\rho(g) \enspace.
\]
Note that $\phi$ takes values in $\C$ while $\rho$ is matrix-valued.  It is a 
fact that a finite group has a finite number of distinct irreducible 
representations (up to isomorphism), and the \emph{Fourier transform} of a 
function $\phi: G \to \C$ is the collection of matrices $\hat{\phi}(\rho)$, 
taken over all distinct irreducible representations $\rho$.

Fixing a group $G$ and a subgroup $H$, we shall focus primarily on the functions 
$\varphi_{c}: G \to \C$ of form
$$
\varphi_{c}(g) = \begin{cases} \frac{1}{\sqrt{|H|}} & \text{if}\;g \in cH,\\
  0 & \text{otherwise,}
\end{cases}
$$
corresponding to the first register of the state $\psi_3$ resulting from Step 3 
above, which is a uniform superposition over the coset $cH$.  The Fourier 
transform of such a function is then
\[
\widehat{\varphi_{c}}(\rho) = \sqrt{\frac{d_\rho}{|G||H|}} \,\rho(c) \cdot 
\sum_{h \in H} \rho(h)\enspace.
\]
Note, as above, that $\widehat{\varphi_{c}}(\rho)$ is a $d_\rho \times d_\rho$ 
matrix.

For any subgroup $H$, the sum $\sum_h \rho(h)$ is precisely $|H|$
times a projection operator (see, e.g., \cite{HallgrenRT00}); we write
$$
\sum_h \rho(h) = |H| \,\pi_H(\rho) \enspace.
$$
With this notation, we can express $\widehat{\varphi_{c}}(\rho)$ as 
$\sqrt{n_\rho}
\,\rho(c) \cdot \pi_H(\rho)$ where $n_\rho = d_\rho |H|/|G|$. For a $d \times d$ 
matrix $M$,
we let $\norm{M}$ denote the matrix norm given by
$$
\norm{M}^2 = \tr \left( M^\dag{} M \right) = \sum_{ij} \abs{M_{ij}}^2,
$$
where $M^{\dag}$ denotes the conjugate transpose of $M$. Then the
probability that we observe the representation $\rho$ is
\begin{align*}
\norm{\widehat{\varphi_c}(\rho)}^2 &= \norm{\sqrt{n_\rho} \,\rho(c) 
\,\pi_H(\rho)}^2\\
&= n_\rho \norm{\pi_H(\rho)}^2 \\
&= n_\rho \,\rank \pi_H(\rho)\enspace,
\end{align*}
where $\rank \pi_H(\rho)$ denotes the rank of the projection operator 
$\pi_H(\rho)$.
See~\cite{HallgrenRT00} for more discussion.

\subsection{Weak vs.\ strong sampling and the choice of basis}

Hallgren, Russell, and Ta-Shma~\cite{HallgrenRT00} show that by measuring only 
the \emph{names} of representations---the so-called \emph{weak standard method} 
in the terminology of~\cite{GrigniSVV01}---it is possible to reconstruct normal 
subgroups (and thus solve the HSP for \emph{Hamiltonian groups}, all of whose 
subgroups are normal).  More generally, this method reconstructs the 
\emph{normal core} of a subgroup, i.e., the intersection of all its conjugates.  
On the other hand, they show that this is insufficient to solve Graph 
Automorphism, since even in an information-theoretic sense this method cannot 
distinguish between the trivial subgroup of $S_n$ and subgroups of order 2 consisting 
of the identity and an involution.

Therefore, in order to solve the HSP for nonabelian groups, we need to measure 
not just the name of the representation we are in, but also the row and column.  
In order for this measurement to be well-defined, we need to choose a basis for 
$U(d_\rho)$ for each $\rho$.  Grigni, Schulman, Vazirani and 
Vazirani~\cite{GrigniSVV01} call this the \emph{strong standard method}.  They 
show that if we measure using a uniformly \emph{random} basis, then trivial and 
non-trivial subgroups are still information-theoretically indistinguishable.  
However, they leave open the question of whether the strong standard method with 
a clever choice of basis, rather than a random one, allows us to solve the HSP 
in nonabelian groups, yielding an algorithm for Graph Automorphism.  

Indeed, in representation theory certain bases are ``preferred'', and have very 
special computational properties, because they give the matrices $\rho(g)$ a 
highly structured or sparse form.  In particular, Moore, Rockmore and 
Russell~\cite{MRR03} showed that so-called \emph{adapted bases} yield highly 
efficient algorithms for the quantum Fourier transform.  


\subsection{Contributions of this paper}

As stated above, \cite{HallgrenRT00} and~\cite{GrigniSVV01} leave an important 
open question: namely, whether there are cases where the \emph{strong standard 
method}, with the proper choice of basis, offers an advantage over a simple 
abelian transform or the \emph{weak standard method}.  We settle this question 
in the affirmative.  Our results deal primarily with the \emph{$q$-hedral} 
groups, i.e., semidirect products of the form $\Z_q \ltimes \Z_p$ where $q \mid 
(p-1)$, and in particular the \emph{affine} groups $A_p \cong \Z_p^* \ltimes 
\Z_p$.

We begin in Section~\ref{sec:full-reconstruction} by focusing on full 
reconstructibility. We define the \emph{Hidden Conjugate Problem} (HCP) as 
follows: given a group $G$, a non-normal subgroup $H$, and a function which is 
promised to be constant on the cosets of some conjugate $bHb^{-1}$ of $H$ (and 
distinct on distinct cosets), determine the subgroup $bHb^{-1}$ by finding an 
element $c \in G$ so that $cHc^{-1} = bHb^{-1}$. We adopt the above 
classification (fully, information-theoretically, quantum information-
theoretically) for this problem in the natural way. Then we show that given a 
subgroup of sufficiently small (but still exponentially large) index, hidden 
conjugates in $A_p$ are fully reconstructible (Theorem~\ref{thm:hcp}).  This 
almost immediately implies that, for prime $q = (p-1)/\polylog(p)$, subgroups of 
the $q$-hedral groups $\Z_q \ltimes Z_p$ are fully reconstructible 
(Theorem~\ref{thm:hsp}).

Section~\ref{sec:info} concerns itself with information-theoretic 
reconstructibility.  We generalize the results of Ettinger and H{\o}yer on the 
dihedral group and show that hidden conjugates of any subgroup are information-
theoretically reconstructible in the affine groups, and more generally the $q$-
hedral groups for all $q$ (Theorem~\ref{thm:infohcp}).  We then show that we can 
identify the order, and thus the conjugacy class, of a hidden subgroup, and this 
implies that all subgroups of the affine and $q$-hedral groups are information-
theoretically reconstructible (Theorem~\ref{thm:infohsp}). 

The results of Sections~\ref{sec:full-reconstruction} and~\ref{sec:info} rely 
crucially on measuring the high-dimensional representations of the affine and 
$q$-hedral groups in a well-chosen basis, namely an \emph{adapted} basis that 
respects the group's subgroup structure.  We show in Section~\ref{sec:random} 
that we lose information-theoretic reconstructibility if we measure using a 
\emph{random} basis instead.  Specifically, we need an exponential number of 
measurements to distinguish conjugates of small subgroups of $A_p$.  This 
establishes for the first time that the strong standard method is indeed 
stronger than measuring in a random basis: some bases provide much more 
information about the hidden subgroup than others.

For some nonabelian groups, the HSP can be solved with a ``forgetful''
approach, where we erase the group's nonabelian structure and perform
an abelian Fourier transform instead.  In Section~\ref{sec:abelian} we
show that this is not the case for the affine groups: specifically, if
we treat $A_p$ as a direct product rather than a semidirect one, its
conjugate subgroups become indistinguishable.

As an application, in Section~\ref{sec:shift} we consider \emph{hidden
  shift} problems. In the setting we consider, one must reconstruct a
``hidden shift'' $s \in \Z_p$ from an oracle $f_s(x)=f(x-s)$, where $f$ is
any function that is constant on the (multiplicative) cosets of a
known multiplicative subgroup of $\Z_p^*$. These functions have been
studied in some depth for their pseudorandom properties, and several
instances have been suggested as cryptographically strong pseudorandom
generators. By associating $f_s$ with its isotropy subgroup, and using
our reconstruction algorithm to find that subgroup, we give an
efficient quantum algorithm for the hidden shift problem in the case
where $f(x)$ is a function of $x$'s multiplicative order mod $r$ for
some $r=\polylog(p)$.  This generalizes the work of van Dam, Hallgren,
and Ip~\cite{vanDamHI03}, who give an algorithm for hidden shift
problems in the case where $f$ is precisely a multiplicative character.

Finally, in Section~\ref{sec:closure} we show that the set of
groups for which the HSP can be solved in polynomial time has the
following closure property: if ${\mathcal H} = \{ H_n \}$ is a family of
groups for which we can efficiently solve the HSP and ${\mathcal K} =
\{ K_n \}$ is a family of groups for which $|K_n| = \polylog | H_n |$,
we can also efficiently solve the HSP for the family $\{ G_n \}$, where
each $G_n$ is any extension of $K_n$ by $H_n$.
This subsumes the results of~\cite{HallgrenRT00} on Hamiltonian
groups, and also those of~\cite{IvanyosMS01} on groups with commutator
subgroups of polynomial size.

\section{The affine and $q$-hedral groups}


Let $A_p$ be the \emph{affine group}, consisting of ordered pairs $(a,b) \in 
\Z_p^* \times \Z_p$, where $p$ is prime, under the multiplication rule 
$(a_1,b_1) \cdot (a_2, b_2) = (a_1a_2, b_1 + a_1b_2)$.  $A_p$ can be viewed as 
the set of affine functions $f_{(a,b)} : \Z_p \to \Z_p$ given by $f_{(a,b)} : x 
\mapsto ax + b$ where multiplication in $A_p$ is given by function composition. 
Structurally, $A_p$ is a semidirect product $\Z_p^* \ltimes \Z_p \cong \Z_{p-1} 
\ltimes \Z_p$.  Its subgroups are as follows:
\begin{itemize} 
\item Let $N \cong \Z_p$ be the normal subgroup of size $p$ consisting of 
elements of the form $(1,b)$.
\item Let $H \cong \Z_p^* \cong \Z_{p-1}$ be the non-normal subgroup of size $p-
1$ consisting of the elements of the form $(a,0)$.  Its conjugates $H^b = (1,b) 
\cdot H \cdot (1,-b)$ consist of elements of the form $(a,(1-a)b)$.  In the 
action on $\Z_p$, $H^b$ is the stabilizer of $b$.
\item More generally, if $a \in \Z_p^*$ has order $q$, let $N_q \cong \Z_q 
\ltimes \Z_p$ be the normal subgroup consisting of all elements of the form 
$(a^t,b)$, and let $H_a$ be the non-normal subgroup $H_a = \langle (a,0) 
\rangle$ of size $q$.  Then $H_a$ consists of the elements of the form $(a^t,0)$ 
and its conjugates $H_a^b=(1,b) \cdot H_a \cdot (1,-b)$ consist of the elements 
of the form $(a^t,(1-a^t)b)$.
\end{itemize}

Construction of the representations of $A_p$ requires that we fix a generator 
$\gamma$ of $\Z_p^*$.  Define $\log: \Z_p^* \to \Z_{p-1}$ to be the isomorphism 
$\log \gamma^t = t$. Let $\omega_p$ denote the $p$th root of unity $\e^{2 \pi 
i/p}$.  Then $A_p$ has $p-1$ one-dimensional representations $\sigma_s$, which 
are simply the representations of $\Z_p^* \cong \Z_{p-1}$, given by 
$\sigma_t((a,b)) = \omega_{p-1}^{t \log a}$.  Moreover, it has one $(p-1)$-
dimensional representation $\rho$ given by 
\begin{equation}
\label{eq:adaptedbasis}
 \rho((a,b))_{j,k} = \left\{ \begin{array}{ll}
   \omega_p^{bj}  & k = aj \bmod p \\
   0              & \mbox{otherwise}
   \end{array} \right. ,
\; 1 \leq j,k < p
\enspace ,
\end{equation}
where the indices $i$ and $j$ are elements of $\Z_p^*$.  
See~\cite[\S8.2]{Serre77} for a more detailed discussion.

Similarly, given prime $p$ and $q \mid p-1$, we consider the \emph{$q$-hedral 
groups}, namely  semidirect products $\Z_q \ltimes \Z_p$.  These embed in $A_p$ 
a natural way: namely, as the normal subgroups $N_q$ defined above.  The 
\emph{dihedral} groups are the special case where $q=2$.

The representations of $\Z_q \ltimes \Z_p$ include the $q$ one-dimensional 
representations of $\Z_q$ given by $\sigma_\ell((a^t,b)) = \omega_q^{\ell t}$ 
for $\ell \in \Z_q$, and $(p-1)/q$ distinct $q$-dimensional representations 
$\rho_k$ given by
\[ \rho_k((a^u,b))_{s,t} = \left\{ \begin{array}{ll}
\omega_p^{k a^s b}  &  t = s + u \bmod q \\
0                     &  \mbox{otherwise}
  \end{array} \right.\enspace ,
\]
for each $0 \leq s,t < q$. Here $k$ ranges over the elements of $\Z_p^* / \Z_q$, 
or, to put it differently, $k$ takes values in $\Z_p^*$ but $\rho_k$ and 
$\rho_{k'}$ are equivalent if $k$ and $k'$ are in the same coset of $\langle a 
\rangle$.  

The representations of the affine and $q$-hedral groups are related as follows.  
The restriction of the $(p-1)$-dimensional representation $\rho$ of $A_p$ to 
$N_q$ is reducible, and is isomorphic to the direct product of the $\rho_k$.  
Moreover, if we measure $\rho$ in a \emph{Gel'fand-Tsetlin} basis such 
as~\eqref{eq:adaptedbasis} which is \emph{adapted} to the tower of subgroups 
\[ A_p > N_q > \Z_p > \{1\} \enspace , \]  
then $\rho$ becomes block-diagonal, with $(p-1)/q$ blocks of size $q$, and these 
blocks are exactly the representations $\rho_k$ of $N_q$.  (See~\cite{MRR03} for 
an introduction to adapted bases and their uses in quantum computation.)  We 
will use this fact in Sections~\ref{sec:info} and~\ref{sec:random} below.

The affine and $q$-hedral groups are \emph{metacyclic} groups, i.e., extensions 
of a cyclic group $\Z_p$ by a cyclic group $\Z_q$.  In~\cite{Hoyer97}, H{\o}yer 
shows how to perform the nonabelian Fourier transform over such groups (up to an 
overall phase factor) with a polynomial, i.e., $\polylog(p)$,  number of 
elementary quantum operations.

\section{Full reconstructibility} 
\label{sec:full-reconstruction}

In this section we show that conjugates of sufficiently large subgroups of the 
affine groups are fully reconstructible in polynomial time.  For some values of 
$p$ and $q$, this allows us to completely solve the Hidden Subgroup Problem for 
the $q$-hedral group $\Z_q \ltimes \Z_p$. 

\begin{theorem} \label{thm:hcp}
Let $p$ be prime and let $a \in \Z_p^*$ have order $q = (p-1) / \polylog(p)$.  
Then the hidden conjugates of $H_a$ in $A_p$ are fully reconstructible.
\end{theorem}

\begin{proof}
Consider first the maximal non-normal subgroup $H = H_\gamma$ (where $\gamma$ is 
a generator of $\Z_p^*$). Carrying out steps 1 through 3 of the Fourier sampling 
procedure outlined in the introduction results in a state $\psi_3$ over the 
group $G$ which is uniformly supported on a random left coset of the conjugate 
$H^b = bHb^{-1}$.  Using the procedure of~\cite{Hoyer97}, we now compute the 
quantum Fourier transform of this state over $A_p$, in the 
basis~\eqref{eq:adaptedbasis}.  The associated projection operator is
$$
\pi_{H^b}(\rho)_{j,k} = \frac{1}{p-1} \;\omega_p^{b(j-k)} \enspace,
$$
for $1 \leq j,k < p$.  This is a circulant matrix of rank one.  More 
specifically, every column is some root of unity times the vector
$$
(u_b)_j = \frac{1}{p-1} \;\omega_p^{bj} \enspace,
$$
$1 \leq j < p$. This is also true of $\rho(c) \cdot \pi_{H^b}(\rho)$; since 
$\rho(c)$ has one nonzero entry per column, left multiplying by $\rho(c)$ simply 
multiplies each column of $\pi_{H^b}(\rho)$ by a phase.  Note that in this case
$$
n_\rho = d_\rho |H|/|G| = (p-1)/p = 1-1/p \enspace,
$$
so that upon measurement the $(p-1)$-dimensional representation $\rho$ is 
observed with overwhelming probability $1 - 1/p$. 

Assuming that we observe $\rho$, we perform another change of basis: namely, we 
Fourier transform each column by left-multiplying $\rho(cH)$ by $Q_{\ell,j} = 
(1/\sqrt{p-1})\;\omega_{p-1}^{-\ell j}$.  In terms of quantum operations, we are 
applying the quantum Fourier transform over $\Z_{p-1}$ to the row register, 
while leaving the column register unchanged.  We can now infer $b$ by measuring 
the frequency $\ell$.  Specifically, we observe a given value of $\ell$ with 
probability
\begin{equation}
  P(\ell)
  = \left| \frac{1}{p-1} \sum_{j=1}^{p-1} \omega_p^{bj} \omega_{p-1}^{-\ell j} 
\right|^2
= \frac{1}{(p-1)^2} \left| \sum_{j=1}^{p-1} \e^{2 i \theta j} \right|^2 
 = \frac{1}{(p-1)^2} \frac{\sin^2 (p-1) \theta}{\sin^2 \theta}
\end{equation}
where 
\[ \theta = \left( \frac{b}{p} - \frac{\ell}{p-1} \right) \pi \enspace. \]
Now note that for any $b$ there is an $\ell$ such that $|\theta| \leq \pi/(2(p-
1))$.  Since
$$
(2x/\pi)^2 \leq \sin^2 x \leq x^2
$$
for $|x| \leq \pi/2$, this gives $P(\ell) \geq (2/\pi)^2$.

Recall that the probability that we observed the $(p-1)$-dimensional 
representation $\rho$ in the first place is $n_\rho = 1-1/p$. Thus if we measure 
$\rho$, the column, and then $\ell$ and then guess that $b$ minimizes 
$|\theta|$, we will be right $\Omega(1)$ of the time.  This can be boosted to 
high probability, i.e., $1-o(1)$, by repeating the experiment a polynomial 
number of times.

Consider now the more general case, when the hidden subgroup is a conjugate of 
the subgroup $H_a$ where $a$'s order $q$ is a proper divisor of $p-1$.  Recall 
that a given conjugate of $H_a$ consists of the elements of the form $(a^t,(1-
a^t)b)$.  Then we have
\[
\pi_{H_a^b}(\rho)_{j,k} = \frac{1}{q} \left\{ \begin{array}{ll}
    \omega_p^{b(j-k)}  & k = a^t j \mbox{ for some } t \\
    0 & \mbox{otherwise}
  \end{array} \right.
\enspace,
\]
for $1 \leq j,k < p$.
In other words, the nonzero entries are those for which $j$ and $k$ lie in the 
same coset of $\langle a \rangle \subset \Z_p^*$.  The rank of this projection 
operator is thus the number of cosets, which is the index $(p-1)/q$ of $\langle 
a \rangle$ in $\Z_p^*$.  Since $n_\rho$ is now $q/p$, we again observe $\rho$ 
with probability 
$$
n_\rho \,\rank \pi_{H_a}(\rho) = (p-1)/p = 1-1/p \enspace.
$$

Following the same procedure as before, we carry out a partial measurement on 
the columns of $\rho$, and then Fourier transform the rows.  After changing the 
variable of summation from $t$ to $-t$ and adding a phase shift of $\e^{-i 
\theta (p-1)}$ inside the $|\cdot|^2$, the probability we observe a frequency 
$\ell$, assuming we find ourselves in the $k$th column, is
\begin{equation}
 \begin{split}
  \label{eq:other}
    P(\ell) & = 
    \left| \frac{1}{\sqrt{q(p-1)}} 
      \,\sum_{t=0}^{q-1} \omega_p^{b(a^t k \bmod p)} \omega_{p-1}^{-\ell (a^t k 
\bmod p)}
    \right|^2 
    \\
    & =  \frac{1}{q(p-1)} \left| 
      \sum_{t=0}^{q-1} \e^{2 i \theta (a^t k \bmod p)} 
    \right|^2
    \enspace.
 \end{split}
\end{equation}
Now note that the terms in the sum are of the form $\e^{i \phi}$ where (assuming 
w.l.o.g.\ that $\theta$ is positive)
$$
\phi \in [-\theta (p-1),\theta (p-1)]\enspace.
$$
If we again take $\ell$ so that $|\theta| \leq \pi/(2(p-1))$, then $\phi \in [-
\pi/2,\pi/2]$ and all the terms in the sum have nonnegative real parts.  We will 
obtain a lower bound on the real part of the sum by showing that a constant 
fraction of the terms have $\phi \in (-\pi/3,\pi/3)$, and thus have real part 
more than $1/2$.  This is the case whenever $a^t k \in (p/6,5p/6)$, so it is 
sufficient to prove the following lemma:

\begin{lemma}
Let $a$ have order $q = p/\polylog(p)$ in $\Z_p^*$, $p$ a prime.
Then at least $(1/3 - o(1)) q$ of the elements in
the coset $\langle a \rangle k$ are in the interval $(p/6,5p/6)$.
\end{lemma}
\noindent
\begin{proof}
We will prove this using \emph{Gauss sums}, which quantify the
interplay between the characters of $\Z_p$ and the characters of
$\Z_p^*$. In particular, Gauss sums establish bounds on the
distribution of powers of $a$.  Specifically, if $a$ has order $q$
in $\Z_p^*$ then for any integer $k \not\equiv 0 \bmod p$ we have
$$
\sum_{t = 0}^{q - 1} \omega_p^{a^t k} = \ord(p^{1/2}) = o(p) \enspace .
$$
(See \cite{KonyaginS99} and Appendix~\ref{appendix:gauss-sums}.)

Now suppose $s$ of the elements $x$ in $\langle a \rangle k$ are in the set 
$(p/6,5p/6)$, for which $\re \omega_p^x \geq -1$, and the other $q-s$ elements 
are in $[0,p/6] \cup [5p/6,p)$, for which $\re \omega_p^x \geq 1/2$.  Thus we 
have 
$$
\re \sum_{t = 0}^{q-1} \omega_p^{a^t k} \geq \,(q/2) - \,(3s/2).
$$
If $s \leq (1/3-\eps) q$ for any $\eps > 0$ this is $\Theta(q)$, a
contradiction.
\end{proof}

Now that we know that a fraction $1/3-\eps$ of the terms in~\eqref{eq:other} 
have real part at least $1/2$ and the others have real part at least $0$, we can 
take $\eps = 1/12$ (say) and write
\[
P(\ell) \geq \frac{1}{q(p-1)} \left( \frac{q}{8} \right)^2 = \frac{1}{64}
\frac{q}{p-1} = \frac{1}{\polylog(p)} \enspace.
\]
Thus we observe the correct frequency with at least polynomially small 
probability; again this can be boosted to high probability by repetition.
\end{proof}

Theorem~\ref{thm:hcp} implies that we can completely solve the Hidden Subgroup 
Problem for certain $q$-hedral groups.

\begin{theorem}
\label{thm:hsp}
Let $p$ and $q$ be prime with $q = (p-1)/\polylog(p)$.  Then subgroups of the 
$q$-hedral group $\Z_q \ltimes \Z_p$ are fully reconstructible.
\end{theorem}

\begin{proof}
First, note that we can fully reconstruct $H$ if it is non-trivial and normal.  
We do this by reconstructing the normal core of $H$,
\[
C(H) = \bigcap_{\gamma \in  G} \gamma H \gamma^{-1}
\]
using the techniques of~\cite{HallgrenRT00} (the weak standard method).  The 
$q$-hedral groups have the special property that no non-normal subgroup contains 
a non-trivial normal subgroup;
then $B$ is normal; 
in particular, if $H$ is non-normal, then $C(H)$ is the trivial subgroup.  Thus 
by reconstructing $C(H)$, we either learn $H=C(H)$ or learn that $H$ is either 
trivial or non-normal.  Furthermore, if $H$ is trivial we will learn this by 
checking our reconstruction against the oracle $f$ and finding that it is 
incorrect.  Therefore, it suffices to consider the non-normal subgroups.

If $q$ is prime, then the non-normal subgroups of $\Z_q \ltimes \Z_p$ are all 
conjugate to a single subgroup $K \cong \Z_q$, so the hidden subgroup problem 
reduces to the hidden conjugate problem for $K$.  While one can construct a 
proof similar to that of Theorem~\ref{thm:hcp} directly for the $q$-hedral 
groups, it is convenient to embed them in $A_p$ using the isomorphisms $N_q 
\cong \Z_q \ltimes \Z_p$ and $H_a \cong K$ and appeal to Theorem~\ref{thm:hcp}.

Now suppose we have an oracle $f: \Z_q \times \Z_p \to S$.  We extend this to an 
oracle $f'$ on $A_p$ as follows.  Choose a generator $\gamma \in \Z_p^*$ and one 
of the $q-1$ elements $a \in \Z_p^*$ of order $q$, and let
\[ f': A_p \to S \times \langle a \rangle \]
where
\[ f'((a,b)) = \left( f\left(\Bigr( \Bigl\lfloor \frac{\log a}{(p-1)/q} \Bigr\rfloor, b\Bigr)\right) , a^q 
\right) \]
recalling that $\log \gamma^t = t$.  The second component of $f'$ serves to 
distinguish the cosets of $N_q$ from each other, while the first component maps 
each coset of $N_q$ to $\Z_q \ltimes \Z_p$ with the element of $\Z_q$ written 
additively, rather than multiplicatively.  (This last step is not strictly 
necessary---after all, we could have written the elements of $A_p$ in additive 
form in the first place---but it can be carried out with Shor's algorithm for 
the discrete logarithm~\cite{Shor97}.)  This reduces the HCP for $K$ (and 
therefore the HSP) on $\Z_q \ltimes \Z_p$ to the HCP for $H_a$ on $A_p$, 
completing the proof.
\end{proof}

As an example of Theorem~\ref{thm:hsp}, if $q$ is a \emph{Sophie Germain} prime, 
i.e., one for which $p=2q+1$ is also a prime, we can completely solve the HSP 
for $\Z_q \ltimes \Z_p$.

\section{Information-theoretic reconstructibility}
\label{sec:info}

In this section, we show that \emph{all} subgroups of the affine and $q$-hedral 
groups, regardless of their size, are information-theoretically reconstructible.  
We start by considering the hidden conjugate problem for subgroups $H_a = 
\langle (a, 0) \rangle$ in $A_p$.  Then in Theorem~\ref{thm:infohsp} we show 
that we can identify the conjugacy class of a  hidden subgroup, and therefore 
the subgroup itself.  This generalizes the results of Ettinger and 
H{\o}yer~\cite{EttingerH98} who show information-theoretic reconstructibility 
for the dihedral groups, i.e., the case $q=2$.

\begin{theorem}
\label{thm:infohcp}
Let $p$ be prime and let $a$ be any element of $\Z_p^*$.  Then the hidden 
conjugates of $H_a$ in $A_p$ are information-theoretically reconstructible.  
\end{theorem}

\begin{proof}
Suppose $a$ has order $q$.  Recall that $H_a$ and its conjugates $H_a^b$ are 
maximal in the subgroup $N_q \cong \Z_q \ltimes \Z_p$.  We wish to show that 
there is a measurement whose outcomes, given two distinct values of $b$, have 
large, i.e., $1/\polylog(p)$, total variation distance.  First, we perform a 
series of partial measurements as follows.  
\begin{itemize}
\item[(i.)] Measure the name of the representation of $A_p$.  If this is not 
$\rho$ try again.  Otherwise, continue; 
\item[(ii.)] Measure the name of the representation $\rho_k$ of $N_q$ inside 
$\rho$; 
\item[(iii.)] Measure the column of $\rho_k$; and 
\item[(iv.)] Perform a POVM with $q$ outcomes, in each of which $s$ is $u$ or 
$u+1 \bmod q$ for some $u \in \Z_q$.
\end{itemize}
As in Theorem~\ref{thm:hcp}, we measure the $(p-1)$-dimensional representation 
of $A_p$ in a chosen basis.  Recall that in the adapted 
basis~\eqref{eq:adaptedbasis} the restriction of $\rho$ to $N_q$ is block 
diagonal, where the $(p-1)/q$ blocks are the $q$-dimensional representations 
$\rho_k$ of $N_q$.  Therefore, the projection operator $\pi_{H_a^b}(\rho)$ is 
block-diagonal, and each of its blocks is one of the projection operators 
$\pi_{H_a^b}(\rho_k)$.  Summing $\rho_k$ over $H_a^b = \{(a^t,(1-a^t)b)\}$ gives 
$$
\left(\pi_{H_a^b}(\rho_k)\right)_{s,t} = (1/q) \; \omega_p^{k(a^s-a^t)b}
$$
for $0 \leq s,t < q$.  This is a matrix of rank 1, where each column (even after 
left multiplication by $\rho_k(c)$) is some root of unity times the vector 
$(u_k)_s = (1/q) \;\omega_p^{k a^s b}$.  Since $n_\rho = q/p$, the probability 
that we observe a particular $\rho_k$ is $q/p$.  Since $\pi_{H_a^b}(\rho)$ has 
$(p-1)/q$ blocks of this kind, it has rank $(p-1)/q$, and the total probability 
that we observe $\rho$ is $(p-1)/p=1-1/p$ as before.

Then these four partial measurements determine $k$, remove the effect of the 
coset, and determine that $s$ has one of two values, $u$ or $u+1$.  Up to an 
overall phase we can write this as a two-dimensional vector
\[ \frac{1}{\sqrt{2}}
\ve
\omega_p^{k a^u b} \\
\omega_p^{k a^{u+1} b} \ctor\enspace.
\]
We now apply the Hadamard transform 
$$
\frac{1}{\sqrt{2}} {\mat 1 & 1 \\ 1 & -1 \rix}
$$
and measure $s$.  The probability we observe that $s=u$ or $u+1$ is then $\cos^2 
\theta$ and $\sin^2 \theta$ respectively, where $\theta = (k a^u (a-1) b 
\pi)/p$.  Now when we observe a $q$-dimensional representation, the $k$ we 
observe is uniformly distributed over $\Z_p^* / \Z_q$, and when we perform the 
POVM, the $u$ we observe is uniformly distributed over $\Z_q$.  It follows that 
the coefficient $m = k a^u (u-1)$ is uniformly distributed over $\Z_p^*$.  For 
any two distinct $b$, $b'$, the total variation distance is then
\[
\frac{1}{2(p-1)} \sum_{m \in \Z_p^*} \left( \left| \cos^2 \frac{\pi m b}{p} - 
\cos^2 \frac{\pi m b'}{p} \right| + 
\left| \sin^2 \frac{\pi m b}{p} - \sin^2 \frac{\pi m b'}{p} \right| \right) 
\enspace .
\]
This we rewrite
\begin{eqnarray*}
& &  \frac{1}{p-1} \sum_{m \in \Z_p^*} \left| \cos^2 \frac{\pi m b}{p} - \cos^2 
\frac{\pi m b'}{p} \right|\\
&  = & \frac{1}{2(p-1)} \sum_{m \in \Z_p} \left| \cos \frac{2 \pi m b}{p} - \cos 
\frac{2 \pi m b'}{p} \right| \\
& \geq &\frac{1}{4(p-1)} \sum_{m \in \Z_p} \left( \cos \frac{2 \pi m b}{p} - 
\cos \frac{2 \pi m b'}{p} \right)^2 \\
&  = & \frac{p}{4(p-1)} > \frac{1}{4} \enspace.
\end{eqnarray*}
(Adding the $m=0$ term contributes zero to the sum in the second line.  In the 
third line we use the facts that $|x| \leq x^2/2$ for all $|x| \leq 2$, the 
average of $\cos^2 x$ is $1/2$, and the two cosines have zero inner product.)

Since the total variation distance between any two distinct conjugates is 
bounded below by a constant, we can 
distinguish between the $p$ different conjugates with only $\ord(\log p) = 
\poly(n)$ samples.  Thus, hidden conjugates in $A_p$ are information-
theoretically reconstructible, completing the proof.
\end{proof}

\smallskip
By embedding the $q$-hedral groups in $A_p$ as in Theorem~\ref{thm:hsp}, we can 
generalize Theorem~\ref{thm:infohcp} to the $q$-hedral groups (note that we do 
not require here that $q$ is prime):
\begin{theorem}
\label{thm:infohcpq}
Let $p$ be prime and $q$ a divisor of $p-1$.  The subgroups of the $q$-hedral 
groups $\Z_q \ltimes \Z_p$ are information-theoretically reconstructible. 
\end{theorem}

We now wish to information-theoretically reconstruct all subgroups of the affine 
and $q$-hedral groups.  We can do this by first reconstructing which conjugacy 
class they lie in, and then applying Theorems~\ref{thm:infohcp} 
and~\ref{thm:infohcpq}.  

\begin{theorem}
\label{thm:infohsp} Let $p$ be prime and $q$ a divisor of $p-1$.  The subgroups 
of the $q$-hedral groups $\Z_q \ltimes \Z_p$ are information-theoretically 
reconstructible. In particular, the subgroups of the affine groups $A_p = 
\Z_{p}^* \ltimes \Z_p$ are information-theoretically reconstructible.
\end{theorem}

\begin{proof}
As in Theorem~\ref{thm:hsp}, we can (fully) reconstruct the normal subgroups of 
$\Z_q \ltimes \Z_p$, so it suffices to consider non-normal subgroups $H$.  
Recall that in this case, $H$ is cyclic and $|H|$ is equal to the order of $a$, 
where $H = \langle(a,b)\rangle$.  Since there is a unique conjugacy class of 
subgroups of each order, it suffices to determine $|H|$, at which point the 
subgroup $H$ can be determined by Theorem~\ref{thm:infohcpq}.

Let the oracle be $f: \Z_q \ltimes \Z_p \to S$, and let $p_1^{\alpha_1}\ldots 
p_k^{\alpha_k}$ be the prime factorization of $q$, in which case $k \leq \sum_i 
\alpha_i = \ord(\log q)$.  For each $i \in \{1, \ldots, k\}$ and each $\alpha 
\in \{0, \ldots, \alpha_i \}$, we will determine if $p_i^{\alpha} \mid |H|$, and 
taking the largest such $\alpha$ for each $i$ gives the prime factorization of 
$|H|$.

To do this, for each $i \in [k]$ and $1 \leq \alpha \leq \alpha_i$, let 
$\Upsilon_i^\alpha: \Z_{q} \ltimes \Z_p \to \Z_{q/p_i^{\alpha}}$ be the 
homomorphism given by
$$
\Upsilon_i^\alpha: (a,b) \mapsto a^{p_i^{\alpha}}\enspace.
$$
Then let
$$
A_i^{\alpha_i} = \ker \Upsilon_i^\alpha = \{ \gamma \in \Z_q \ltimes \Z_p \mid 
\gamma^{p_i^{\alpha_i}} = \id \} \enspace,
$$
where $\id$ denotes the identity element of $\Z_q \ltimes \Z_p$. 
$A_i^{\alpha_i}$ is the subgroup of $\Z_q \ltimes \Z_p$ consisting of all 
elements whose orders are a multiple of $p_i^{\alpha}$.  Consider now the 
function
$$
f' : \Z_q \ltimes \Z_p \to S \times \Z_{q/p_i^\alpha}
$$
given by 
\[ 
f'(\gamma) = \left( f(\gamma),\Upsilon_i^\alpha(\gamma) \right) \enspace . 
\] 
Observe that $f'$ is constant (and distinct) on the left cosets of $H \cap 
A_i^{\alpha}$ and, furthermore, the subgroup $H \cap A_i^\alpha$ has order 
$p^\alpha$ if and only if $p^\alpha$ divides the order of $a$. We may then 
determine if $H \cap A_i^\alpha$ has order $p^\alpha$ by assuming that it does, 
reconstructing $H$ with Theorem~\ref{thm:infohcpq} using $f'$ as the oracle, and 
checking the result against the original oracle $f$. This allows us to determine 
the prime factorization of $|H|$ as desired.  Therefore, all subgroups of the 
$q$-hedral groups $\Z_q \ltimes \Z_p$ are information-theoretically 
reconstructible. 
\end{proof}

\smallskip As in the dihedral case~\cite{EttingerH98}, we know of no
polynomial-time algorithm which can reconstruct the most likely $b$
from these queries.  However, Kuperberg~\cite{Kuperberg03} gives a
quantum algorithm for the HSP in the dihedral group, and more
generally the hidden shift problem, that runs in subexponential
($\e^{\ord(\log^{1/2} p)}$) time.  Since we can reduce the HSP on
$\Z_q \ltimes \Z_p$ to a hidden shift problem by focusing on two
cosets of $\Z_p$, this algorithm applies to the $q$-hedral groups as
well.

\section{Random vs.\ adapted bases}
\label{sec:random}

In Theorems~\ref{thm:infohcp} and~\ref{thm:infohsp}, we measured the high-
dimensional representation $\rho$ in a specific basis which is adapted to the 
subgroup structure of $A_p$ and the $q$-hedral groups.  In contrast, we show in 
this section that if we measure $\rho$ in a \emph{random} basis instead, then 
for all but the largest values of $q$ we need an exponential number of 
measurements in order to information-theoretically distinguish conjugate 
subgroups from each other.

\begin{theorem}  
Let $p$ be prime and let $a \in \Z_p^*$ have order $q$ where $q < p^{1-\eps}$ 
for some $\eps > 0$.  Let $P_b(v)$ be the probability that we observe a basis 
vector $v$ in the Fourier basis if the hidden subgroup is $H_a^b$.  If we 
measure $\rho$ in a random basis, then for any two $b, b'$, with high 
probability the $L_1$ distance between these probability distributions is 
exponentially small, i.e., there exists $\beta > 0$ such that
\[ \sum_v \left| P_b(v) - P_{b'}(v) \right| < p^{-\beta} \enspace . \]
Thus it takes an exponentially large number of measurements to distinguish the 
conjugates $H_a^b$ and $H_a^{b'}$.
\end{theorem}

\begin{proof}
Since we observe the high-dimensional representation $\rho$ with probability $1-
1/p$, it suffices to consider the $L_1$ distance summed over the $d_\rho=p-1$ 
basis vectors of $\rho$.  In fact, we will show that $P_b(v)$ is exponentially 
close to the uniform distribution for all $b$.

Write $\pi = \pi_{H_a^b}(\rho)$.  Then the probability we observe a given basis 
vector $v$, conditioned on observing $\rho$, is
\[ P_b(v) = \frac{1}{\rank \pi} \abs{\pi \cdot v}^2 \enspace . \]
If $v$ is uniformly random with norm $1$, the expectation of $\abs{\pi \cdot 
v}_2^2$ is $(\rank \pi)/d_\rho$, and so the expectation of $P_b(v)$ is 
$1/d_\rho$.  We will use the following lemma to show that when $\rank \pi$ is 
sufficiently large, $P_b(v)$ is tightly concentrated around this expectation.

\begin{lemma}
\label{lem:bound}  
Let $\pi$ be a projection operator of rank $r$ in a $d$-dimensional space, and 
let $v$ be a random $d$-dimensional vector of unit length.  Then for all $0 < 
\delta < 2$, 
\[ \Pr\left[ \,\left| \abs{\pi \cdot v}_2^2 - \frac{r}{d} \right| > \delta 
\frac{r}{d} \right] 
< 4 \e^{-r \delta^2 / 48} \enspace . \]
\end{lemma}

\begin{proof}
We use an argument similar to~\cite{GrigniSVV01}.  We can think of a random $d$-
dimensional complex vector $v$ as a random $2d$-dimensional real vector of the 
same length, and we can think of this in turn as 
\[ v_i = \frac{w_i}{\sum_{i=1}^{2d} w_i^2} \]
where the $w_i$ are independent Gaussian variables with zero mean and unit 
variance.  By choosing a basis in which $\pi$ projects onto the first $r$ 
(complex) components of $v$, we have
\[ \abs{\pi \cdot v}_2^2 = \frac{\sum_{i=1}^{2r} w_i^2}{\sum_{i=1}^{2d} w_i^2} 
= \frac{r}{d} \frac{(1/2r) \sum_{i=1}^{2r} w_i^2}{(1/2d) \sum_{i=1}^{2d} w_i^2} 
\enspace . 
\]
We now use the following Chernoff bound, which can be derived from the moment 
generating function.  For any $t$, we have
\[
 \Pr\left[ \,\left| \left( \frac{1}{t} \sum_{i=1}^t w_i^2 \right) - 1 \right| > 
\eps \right]
< 2 \left[ (1+\eps)^{1/2} \,\e^{-\eps/2} \right]^t \enspace . 
\]
For $|\eps| < 1/2$, we have $\ln (1+\eps) < \eps - \eps^2/3$ and this becomes
\begin{equation}
\label{eq:chernoff}
\Pr\left[ \,\left| \left( \frac{1}{t} \sum_{i=1}^t w_i^2 \right) - 1 \right| > 
\eps \right]
< 2 \e^{-t \eps^2 / 6} \enspace . 
\end{equation}

Now, for any $a,b$, if $|a/b - 1| > \delta$ where $\delta < 2$, then either $|a-
1| > \delta/4$ or $|b-1| > \delta/4$.  Taking the union bound over these events 
where $a = (1/2r) \sum_{i=1}^{2r} w_i^2$ and $b = (1/2d) \sum_{i=1}^{2d} w_i^2$, 
setting $\eps = \delta/4$ and $t=2r \leq 2d$ in~\eqref{eq:chernoff} gives the 
stated bound.
\end{proof}

Setting $d=d_\rho$ and $r = \rank \pi$, Lemma~\ref{lem:bound} and the union 
bound imply that, for any constant $A > \sqrt{48}$, if
\begin{equation}
\label{eq:delta}
 \delta = A \sqrt{ \frac{\log d_\rho}{\rank \pi}} 
\end{equation}
then, with high probability, for all $d_\rho$ basis vectors $v$ we have
\[ \abs{ P_b(v) - \frac{1}{d_\rho} } < \frac{\delta}{d_\rho} \enspace . \]
Summing over all $v$, this implies that the $L_1$ distance between $P_b(v)$ and 
the uniform distribution is at most $\delta$.  Now recall that $\rank \pi = (p-
1)/q$.  If $q < p^{1-\eps}$, then $\rank \pi > p^\eps$, and~\eqref{eq:delta} 
gives $\delta < p^{-\beta}$ where $\beta = \eps/3$, say.  Since $P_b(v)$ is 
within $\delta$ of the uniform distribution for all $b$, doubling the constant 
$A$ and using the triangle inequality completes the proof.
\end{proof}

Several remarks are in order.  First, just as for the dihedral group, we can 
information-theoretically distinguish conjugate subgroups if we use a random 
basis \emph{within} each $q$-dimensional block.  The problem is that rather than 
having this block-diagonal structure, a random basis cuts across these blocks, 
mixing different ``frequencies'' $\rho_k$ and canceling out the useful 
information.  This is precisely because it is not adapted to the subgroup 
structure of $A_p$; it doesn't ``know'' that $\rho$ decomposes into a direct sum 
of the $\rho_k$.

Second, it is worth noting that for the values of $q$ for which we have an 
algorithm for full (as opposed to information-theoretic) reconstruction, namely 
$q=p/\polylog(p)$, a random basis works as well since the $L_1$ distance 
$\delta$ becomes $1/\polylog(p)$.  Based on the strong evidence from 
representation theory that some bases are much better for computation than 
others, we conjecture that, for some families of groups, adapted bases allow 
full reconstruction while random bases do not; but this remains an open 
question.

Third, while we focused above on distinguishing conjugate subgroups from each 
other, in fact our proof shows that if $q < p^{1-\eps}$ a random basis is incapable 
of distinguishing $H_a$ from the \emph{trivial} subgroup.  In contrast, 
Theorems~\ref{thm:infohcp} and~\ref{thm:infohsp} show that an adapted basis 
allows us to do this.

\section{Failure of the abelian Fourier transform}
\label{sec:abelian}

In \cite{EttingerH98} the abelian Fourier transform over $\Z_2 \times \Z_p$ is 
used in a reconstruction algorithm for the dihedral groups. Using this sort of 
``forgetful'' abelian Fourier analysis it is similarly information-theoretically 
possible to reconstruct subgroups of the $q$-hedral groups, when $q$ is small 
enough.

However, it does not seem possible to reconstruct subgroups of $A_p$ using the 
abelian Fourier transform.  In particular, we show in this section that if we 
think of the affine group as a direct product $\Z_p^* \times \Z_p$ rather than a 
semidirect product, then the conjugates of the maximal subgroup become 
indistinguishable.  This is not surprising, since in an abelian group conjugates 
are identical by definition, but it helps illustrate that nonabelian hidden 
subgroup problems require nonabelian approaches (most naturally, in
our view, representation theory).

Let us consider the hidden conjugate problem for the maximal subgroup $H$, i.e., 
$H_a$ where $a$ is a generator of $\Z_p^*$.  In that case, the characters of 
$\Z_p^* \times \Z_p$ are simply $\rho_{k,\ell}(a^t,b) = \omega_{p-1}^{kt} 
\omega_p^{\ell b}$.  Summing these over $H_a = \{ (a^t, (1-a^t)b \}$ shows that 
we observe the character $(k,\ell)$ with probability
\begin{align*}
P(k,\ell)  &= \frac{1}{p \,(p-1)^2} \left| \sum_{t \in \Z_{p-1}} \omega_{p-
1}^{kt} \omega_p^{\ell (1-a^t) b} \right|^2 \\
&= \frac{1}{p \,(p-1)^2} \left| 
   \sum_{x \in \Z_p^*} \omega_{p-1}^{k \log_a x} \omega_p^{-\ell x b} 
  \right|^2
\enspace. 
\end{align*}
This is the inner product of a multiplicative character with an additive one, 
which is another Gauss sum.  In particular, assuming $b \neq 0$, we have 
\begin{eqnarray*}
P(0,0) & = & 1/p \\
P(0, \ell \neq 0) & = & 1/ (p\,(p-1)^2) \\
P(k \neq 0, 0) & = & 0 \\
P(k \neq 0, \ell \neq 0) & = & 1/(p-1)^2
\end{eqnarray*}  
(see Appendix~\ref{appendix:gauss-sums}).  Since these probabilities don't 
depend on $b$, the different conjugates $H_a^b$ with $b \neq 0$ are 
indistinguishable from each other.  Thus it appears essential to use the 
nonabelian Fourier transform and the high-dimensional representations of $A_p$.

\section{Hidden shift problems}
\label{sec:shift}

Using the natural action of the affine group on $\Z_p$, we can apply
our algorithm for the hidden conjugate problem studied above to a
natural family of \emph{hidden shift problems}.  Specifically, let $M$
be a multiplicative subgroup of $\Z_p^*$ of index $r > 1$, let $S$ be
some set of $r+1$ symbols, and let $f: \Z_p \to S$ be a function for
which
$$
f(x) = f(mx) \Leftrightarrow m \in M
$$
for every $x \in \Z_p$.  Observe that $f$ is constant on the
(multiplicative) cosets of $M$ and takes distinct values
on distinct cosets; to put it differently, $f(x)$ is an injective function 
of the multiplicative order of $x$ mod $r$.  
Furthermore, $f(0) \neq f(x)$ for any nonzero $x$.
The hidden shift problem associated with $f$ is
the problem of determining an unknown element $s \in \Z_p$ given oracle
access to the shifted function
$$
f_s(x) = f(x - s)\enspace.
$$
Such functions have remarkable pseudorandom properties, and have been
proposed as pseudorandom generators for cryptographic purposes, where 
$s$ acts as the seed\remove{ or secret key} to generate the sequence (e.g.~\cite{damgard}).

The special case when $f: \Z_p \to \C$ is a \emph{Legendre symbol}, that
is, a multiplicative character of $\Z_p^*$ extended to all of $\Z_p$
by setting $f(0) = 0$, was studied by van Dam, Hallgren, and
Ip~\cite{HallgrenIvD}. They give efficient quantum algorithms for
these hidden shift problems for all characters of $\Z_p^*$. Their
algorithms, however, make explicit use of the complex values taken by
the character, whereas the algorithms we present here depend only on
the symmetries of the underlying function $f$; in particular, in our
case $f$ can be an arbitrary injective function from a multiplicative
character into a set $S$.  On the other hand, their algorithms are
efficient for characters of any order, while our algorithms require
that $r$ be at most polylogarithmic in $p$.

Returning to the general problem defined above, let ${\mathcal F}(\Z_p, S)$
denote the collection of $S$-valued functions on $\Z_p$. Note that the
affine group $A_p$ acts on the set ${\mathcal F}(\Z_p,S)$ by assigning $\alpha \cdot g(x) =
g(\alpha^{-1}(x))$ for each $\alpha \in A_p$ and $g \in F(\Z_p,S)$.  In particular, 
$f_s = (1,s) \cdot f$.  

Now note that the isotropy subgroup of $f$, namely the subgroup of $A_p$
that fixes the cosets of $M$, is precisely $H_a = \langle (a,0) \rangle$ where $a
\in \Z_p^*$ has order $q=(p-1)/r$.  As we have $f_s = (1,s) \cdot f$, the isotropy
subgroup of $f_s$ is the conjugate subgroup $H_a^s = (1,s) \cdot H_a \cdot
(1,-s)$. Observe now that if we define $F_s : A_p \to (\Z_p)^p$ so that
$F_s(\alpha)$ is the $p$-tuple $(\alpha f_s(0), \alpha f_s(1), \ldots, \alpha f_s(p-1))$ then
\begin{equation}
\label{eqn:shift-symmetry}
F_s(\alpha) = F_s(\beta) \Leftrightarrow \alpha^{-1} \beta \in H_a^s \enspace ,
\end{equation}
i.e., $F_s$ is constant precisely on the left cosets of $H_a^s$.  
Evidently, then, the solution to the hidden conjugate problem given by the
oracle $F_s$ determines the solution to the hidden shift problem given
by $f_s$.  Unfortunately, the \emph{values} of the oracle $F_s$ are of exponential
size---we cannot afford to evaluate $\alpha f_s(x)$ for all $x \in \Z_p$. 
This same symmetries expressed in Equation~\eqref{eqn:shift-symmetry}, 
however, can be obtained efficiently by
selecting an appropriate subset $R = \{x_1, \ldots, x_m\} \subset \Z_p$ and
considering the oracle that samples $\alpha f_s$ on $R$: that is, 
\[ F^R_s(\alpha) = (\alpha f_s(x_1), \ldots, \alpha f_s(x_m)) \enspace . \] 
Of course, we have $\alpha f_s = \beta f_s \Rightarrow F^R_s(\alpha) = F^R_s(\beta)$
regardless of $R$; the difficulty is finding a small set $R$ for
which $F^R_s(\alpha) = F^R_s(\beta) \Rightarrow \alpha f_s = \beta f_s$. 
We show below that a set of $O(\log p)$ elements selected uniformly at random 
from $\Z_p$ has this property with high probability.





Considering that $\alpha f_s(x) = \alpha \cdot (1,s) \cdot f(x)$, it suffices to show that
if $\alpha f \neq \beta f$ then
\[
\Pr_x[\alpha f(x) = \beta f(x)] \leq 1/2\enspace,
\]
where $x$ is selected uniformly at random in $\Z_p$. Note that for
affine functions $\alpha$ and $\beta$ and an element $x \in \Z_p$ for which
$\beta^{-1}(x) \neq 0$,
$$
\alpha f(x) = \beta f(x) \;\Leftrightarrow\; \frac{\alpha^{-1}(x)}{\beta^{-1}(x)} \in M \enspace .
$$
The function $\alpha^{-1}(x)/\beta^{-1}(x)$ is a \emph{fractional linear
  transform}, i.e., the ratio of two linear functions; these is the
discrete analog of a M\"{o}bius transformation in the complex
plane. As in the complex case, the fractional linear transform $\gamma(x) /
\delta(x)$ is a bijection on the projective space 
$\Z_p \cup \{ \infty \}$ unless $\gamma$ and $\delta$ share a root, 
or, equivalently, there is a scalar $z \in \Z_p^*$ such that
$\gamma(x) = z\delta(x)$.  If $\alpha^{-1}(x) / \beta^{-1}(x)$ is injective, we can
immediately conclude that
$$
\Pr_{x} [ \alpha f(x) = \beta f(x) ] \leq |M|/(p-1) = 1/r \leq 1/2 \enspace.
$$
Otherwise, $\alpha^{-1}(x)/ \beta^{-1}(x) = z$ for some scalar $z$.  
Since $\alpha f \neq \beta f$, however, in this case we must have $z \in \Z_p^* \setminus M$.  
In particular, $f(zy) \neq f(y)$ for any $y \neq 0$, and so
$$
\Pr_{x} [ \alpha f(x) = \beta f(x) ] = 1/p
$$
since this only occurs at the unique root $x$ of $\alpha^{-1}(x)=0$.

In either case, then, $\alpha f$ and $\beta f$ differ on at least half the elements of $\Z_p$ 
whenever $\alpha$ and $\beta$ belong to different cosets of $H_a^s$.  It follows that if
$R \subset \Z_p$ consists of $m$ elements chosen
independently and uniformly at random from $\Z_p$, we have
$$
\Pr_{R} \left[ \forall x \in R, \alpha f(x) = \beta f(x)\right] \leq 1/2^m
$$
for any $\alpha, \beta \in A_p$ with $\alpha^{-1}\beta \notin H_a$. Taking a union bound over
all pairs of left cosets of $H_a$, 
$$
\Pr_{R} \left[ \exists \alpha, \beta \in A_p: \alpha^{-1}\beta \notin H_a, \forall x \in R, \alpha f(x) = \beta
  f(x)\right] \leq \left(\frac{p(p-1)}{|H_a|}\right)^2\frac{1}{2^m}\enspace.
$$
Selecting $m = 5 \log p$ ensures that this probability is less than $1/p$.  

Since we showed in Section~\ref{sec:full-reconstruction} that we can identify a 
hidden conjugate of $H_a$ whenever $H_a$ is of polylogarithmic index in $\Z_p^*$, 
and since this index is $(p-1)/q = r$, this provides an efficient solution to the hidden shift 
problem so long as $r = \polylog(p)$.

\section{Closure under extending small groups}
\label{sec:closure}

In this section we show that for any polynomial-size group $K$ and any $H$ for 
which we can solve the HSP, we can also solve the HSP for any extension of $K$ 
by $H$, i.e., any group $G$ with $K \lhd G$ and $G/K \cong H$.  (Note that this 
is more general than split extensions, i.e., semidirect products $H \ltimes K$.)  
This includes the case discussed in~\cite{HallgrenRT00} of Hamiltonian groups, 
since all such groups are direct products (and hence extensions) by abelian 
groups of the quaternion group $Q_8$~\cite{Rotman94}.  It also includes the case 
discussed in~\cite{FriedlIMSS02} of groups with commutator subgroups of 
polynomial size, such as extra-special $p$-groups, since in that case $K=G'$ and 
$H \cong G/G'$ is abelian.  Indeed, our proof is an easy generalization of that 
in~\cite{FriedlIMSS02}.

\begin{theorem}
\label{thm:semik}
Let $H$ be a group for which hidden subgroups are fully reconstructible, and $K$ 
a group of polynomial size in $\log |H|$.  Then hidden subgroups in any 
extension of $K$ by $H$, i.e., any group $G$ with $K \lhd G$ and $G/K \cong H$, 
are fully reconstructible.
\end{theorem}

\noindent 
\begin{proof}
We assume that $G$ and $K$ are encoded in such a way that multiplication can be 
carried out in classical polynomial time.  We fix some transversal $t(h)$ of the 
left cosets of $K$. First, note that any subgroup $L \subseteq G$ can be 
described in terms of i) its intersection $L \cap K$, ii) its projection $L_H = 
L/(L \cap K) \subseteq H$, and iii) a representative $\eta(h) \in L \cap (t(h) 
\cdot K)$ for each $h \in L_H$.  Then each element of $L_H$ is associated with 
some left coset of $L \cap K$, i.e., $ L = \bigcup_{h \in L_H} \eta(h) \cdot (L 
\cap K)$.  Moreover, if $S$ is a set of generators for $L \cap K$ and $T$ is a 
set of generators for $L_H$, then $S \cup \eta(T)$ is a set of generators for 
$L$.

We can reconstruct $S$ in classical polynomial time simply by querying the 
function $h$ on all of $K$.  Then $L \cap K$ is the set of all $k$ such that 
$f(k) = f(1)$, and we construct $S$ by adding elements of $L \cap K$ to it one 
at a time until they generate all of $L \cap K$. 

To identify $L_H$, as in~\cite{FriedlIMSS02} we define a new function $f'$ on 
$H$ consisting of the unordered collection of the values of $f$ on the 
corresponding left coset of $K$: 
$$
f'(h) = \{ f(g) \mid g \in t(h) \cdot K \}.
$$
Each query to $f'$ consists of $|K| = \poly(n)$ queries to $f$.  The level sets 
of $f'$ are clearly the cosets of $L_H$, so we reconstruct $L_H$ by solving the 
HSP on $H$.  This yields a set $T$ of generators for $L_H$.

It remains to find a representative $\eta(h)$ in $L \cap (t(h) \cdot K)$ for 
each $h \in T$.  We simply query $f(g)$ for all $g \in t(h) \cdot K$, and set 
$\eta(h)$ to any $g$ such that $f(g) = f(1)$.  Since $|T| = \ord(\log |H|) = 
\poly(n)$ this can be done in polynomial time, completing the proof.
\end{proof}

Unfortunately, we cannot iterate this construction more than a constant number 
of times, since doing so would require a superpolynomial number of queries to 
$f$ for each query of $f'$.  If $K$ has superpolynomial size it is not clear how 
to obtain $\eta(h)$, even when $H$ has only two elements.  Indeed, this is 
precisely the difficulty with the dihedral group.

\section{Conclusion and directions for further work}

We have shown that the ``strong standard method,'' applied with
adapted bases, solves in quantum polynomial time certain nonabelian 
Hidden Subgroup Problems that are not solved with any
other known technique, specifically measurements in random
bases or ``forgetful'' abelian approaches.

While we are still very far from an algorithm for HSP in the symmetric group $S_n$ or 
for Graph Automorphism, a global understanding of the power of strong Fourier 
sampling remains an important goal.  Perhaps the next class of groups to try 
beyond the affine and $q$-hedral groups are matrix groups such as ${\rm 
PSL}_2(p)$, whose maximal subgroups are isomorphic to $A_p$, and which include 
one of the infinite families of finite simple groups.

\bigskip {\bf Acknowledgements.}  We are grateful to Wim van Dam,
Julia Kempe, Greg Kuperberg, 
Frederic Magniez, Martin R\"{o}tteler, and Miklos Santha for helpful
conversations, and to Sally Milius and Tracy Conrad for their support.
Support for this work was provided by the California Institute of
Technology's Institute for Quantum Information (IQI), the Mathematical
Sciences Research Institute (MSRI), the Institute for Advanced Study
(IAS), NSF grants ITR-0220070, ITR-0220264, CCR-0093065,
EIA-0218443, QuBIC-0218563, CCR-0049092, 
the Charles Lee Powell Foundation, and the Bell Fund.



\appendix

\section{Notes on exponential sums}
\label{appendix:gauss-sums}

The basic \emph{Gauss sum} bounds the inner products of additive and
multiplicative characters of $\F_p$, the finite field of prime
cardinality $p$. Definitive treatments appear in~\cite[{\S}5]{LidlN97}
and~\cite{KonyaginS99}. Considering $\F_p$ as an additive group with
$p$ elements, we have $p$ additive characters $\chi_s : \F_p \to \C$, for
$s \in \F_p$, given by $\chi_s : z \mapsto \omega_p^{sz}$, where, as above, $\omega_p =
\e^{2 \pi i/p}$ is a primitive $p$th root of unity.  Likewise
considering the elements of $\F_p^* = \F_p \setminus \{ 0\}$ as a multiplicative
group, we have $p-1$ characters $\psi_t : \F_p^* \to \C$, for $t \in \F_p^*$,
given by $\psi_t : g^z \mapsto \omega_{p-1}^{tz},$ where $\omega_{p-1} = \e^{2 \pi i/(p-
  1)}$ is a primitive $(p-1)$th root of unity and $g$ is a
multiplicative generator for the (cyclic) group $\F_p^*$.

With this notation the basic Gauss sum is the following:
\begin{theorem}
  Let $\chi_s$ be an additive character and $\psi_t$ a multiplicative
  character of $\F_p$.  If $s \neq 0$ and $t \neq 1$ then
  $$
  \Bigl | \sum_{z \in \F_p^*} \chi_s(z) \,\psi_t(z) \Bigr | = \sqrt{p}.
  $$
  Otherwise
  $$
  \sum_{z \in \F_p^*} \chi_s(z) \psi_t(z) = \begin{cases}
    p-1 & \text{if}\; s = 0, t = 1,\\
    -1  & \text{if}\; s = 0, t \neq 1,\\
    0   & \text{if}\; s \neq 0, t = 1.\\
  \end{cases}
  $$
\end{theorem}
See~\cite[{\S}5.11]{LidlN97} for a proof.

This basic result has been spectacularly generalized. In the body of
the paper we require bounds on additive characters taken over
multiplicative subgroups of $\F_p^*$. Such sums are discussed in detail
in \cite{KonyaginS99}. The specific bound we require is the following.

\begin{theorem}
  Let $\chi_t$ be a nontrivial additive character of $\F_p$ and $a \in
  \F_p^*$ an element of multiplicative order $q$. Then 
  $$
  \sum_{z = 0}^{q - 1} \chi_t(a^z) = \begin{cases}
    \ord(p^{1/2}), &\text{if}\;q \geq p^{2/3},\\
    \ord(p^{1/4} q^{3/8}), &\text{if}\;p^{1/2} \leq q \leq p^{2/3},\\
    \ord(p^{1/8} q^{5/8}), &\text{if}\;p^{1/3} \leq q \leq p^{1/2}.
  \end{cases}
  $$  
\end{theorem}
See \cite[{\S}2]{KonyaginS99} for a proof. 

Note that in the body of the paper, we use $\Z_p$ to denote the
additive group of integers modulo $p$ and $\Z_p^*$ to denote the
multiplicative group of integers modulo $p$.
\end{document}